\documentclass[journal=jacsat,manuscript=article]{achemso}

\usepackage{xcolor}
\usepackage[mathscr]{euscript}
\usepackage[version=3]{mhchem} 

\author{Sudipta Khamrui}
\affiliation{Department of Physics, Indian Institute of Technology Kharagpur, Kharagpur, 721302, India.}
\author{Kamini Bharti}
\affiliation{Department of Physics, Indian Institute of Technology Kharagpur, Kharagpur, 721302, India.}
\author{Daniella Goldfarb}
\affiliation{Department of Chemical and Biological Physics, Weizmann Institute of Science, Rehovot, 76100, Israel.}
\author{Tilak Das}
\affiliation{Department of Physics, Indian Institute of Technology Kharagpur, Kharagpur, 721302, India.}
\email{tilak.das@phy.iitkgp.ac.in}
\author{Debamalya Banerjee}
\affiliation{Department of Physics, Indian Institute of Technology Kharagpur, Kharagpur, 721302, India.}
\email{debamalya@phy.iitkgp.ac.in}

\title[An \textsf{achemso} demo]
  {Strain-driven Charge Localisation and Spin
Dynamics of Paramagnetic Defects in S-deficit
2H-MoS$_2$ Nanocrystals}

\abbreviations{IR,NMR,UV}

\begin{document}


\begin{abstract}
 A microscopic control over the origin and dynamics of localised spin centres in lower dimensional solids turns out to be a key factor for next generation spintronics and quantum technologies. With the help of low temperature electron paramagnetic resonance (EPR) measurements, supported by the first-principles calculations within density functional theory (DFT) formulation, we found the origin of different high-spin paramagnetic intrinsic charge-centres, Mo$^{3+}$($4d^3$) and Mo$^{2+}$($4d^4$) present in the nano-crystalline sulfur deficit hexagonal molybdenum disulfide (2H-MoS$_{2-x}$), against the established notion of spin-$\frac{1}{2}$, Mo$^{5+}$ centres. A critical strain generated in the nano-structured 2H-MoS$_{2-x}$ was found to be very crucial for spin-localization in this layered material. Indeed, computationally effective proposition of the PBE+$U$ exchange-correlations within DFT including D3-dispersion corrections found to be more viable than expensive higher rung of exchange-correlation functionals, explored earlier. It is also found that the oxygen vacancy of the reduced oxide phase, embedded in 2H-MoS$_{2-x}$ host lattice, has the longest relaxation times. Moreover, the temperature dependence of spin-lattice relaxation measurements reveals a direct process for interstitial spin centres and a Raman process for both sulfur and oxygen vacancy sites. We expect such observation would be a valuable pillar for better understanding of the next generation quantum technologies and device applications.\\

\end{abstract}

\section{Introduction}
The discovery of the two-dimensional (2D) solids stacked by inter-layer van der Waals forces, such as graphene and its nano-structures, group-VI transition metal dichalcogenides (TMDC) as well as the transition metal carbides (MXenes), turn out to be an exciting 2D materials family for their extraordinary electronic, mechanical, optical and spin-properties\cite{Novoselov2004,Zhang2005,Geim2007,Chakraborty2017}. Subsequently, the spin properties of these 2D-materials are rapidly explored in the domain of \textit{spintronic} and \textit{magnetoelectronic} device applications \cite{wolf2001spintronics, PRINZ1999, huang2021room, el2020progress}. In this aspect, defect based spin centres are also explored to verify their applicability in quantum information based  technology \cite{awschalom2013quantum}. Nitroxide-functionalized graphene nano-ribbons, nitrogen vacancy centre of 2D-hexagonal boron nitride and diamond are notable examples in this area \cite{slota2018magnetic, gottscholl2021room, hughes2023two} which show prolonged spin coherence (of afew $\mu s$), even at the room temperature. These localised spin-centres are also the potential sources for single photon emission, which is a recent key interest in the domain of quantum information processing \cite{grosso2017tunable, bathen2021manipulating}.

Apart from these nitrogen based defects, some recent works are also reported on TMDC defect centres, especially on 2D-MoS$_2$, to explore the effectiveness of single photon emission \cite{michaelis2022single}. For example, controlled generation of sulfur vacancies by focused helium ion bombardment on monolayer MoS$_2$ has been reported \cite{barthelmi2020atomistic, mitterreiter2021role}. However, optical coherence of these quantum emitters is limited by many factors such as phonon scattering, nuclear-spin fluctuations and hyperfine interaction \cite{borri2001ultralong, kuhlmann2013charge, utzat2019coherent}. So, a study of spin dynamics of these native defect centres in 2D-MoS$_2$ to verify their applicability in the era of quantum technology applications is desirable. There are some reports of spin dephasing measurements on the resident electrons through time-resolved optical measurements of few TMDCs\cite{jiang2022coherent, yang2015spin, yang2015long}. However, there are still lack of clear understanding in the overall spin dimension of defect centres in MoS$_2$ from microscopic point of view.

The defects in MoS$_2$ are also reported to induce magnetism in the host material, such as defect-induced ferromagnetic ordering \cite{cai2015vacancy, han2016electron, han2016investigation, zheng2014tuning, sanikop2020robust}. A local strain around the sulfur vacancies (single S-vacancy, say S$_v$, multi vacancies, say MoS$_3$ or MoS$_6$ centres) was shown to induce 2H to 1T-phase transformation in MoS$_2$, which led to the ferromagnetic ordering. Nevertheless, there may be several other intrinsic defects present in the samples e.g. anti-sites or interstitials \cite{zhou2013intrinsic, Tsunetomo2021, xu2022frenkel}, which may lead to spin localisation and induce magnetism in the host lattice. In the paramagnetic state, these defects can be probed through electron paramagnetic resonance (EPR) spectroscopy. So far, most of these paramagnetic defects in MoS$_2$ have been considered as spin-$\frac{1}{2}$ (Mo$^{5+}$-charge centres) \cite{zeng2019sulfur, martinez2018paramagnetic, panich2009magnetic, anbalagan2023gamma, xia2019sulfur}. However, this spin state assignment seems inadequate to explain the wide EPR linewidth of some of the MoS$_2$ samples \cite{cai2015vacancy, guo2022charge, sanikop2019tailoring, jena2022defect, tang20201, cai2021ultrahigh, jiao2019defect}, which requires further investigation.

Besides growing potential of defect engineering in the field of spintronics and quantum information technological applications, very limited efforts have been employed so far to understand the spin-localisation and spin-states of the defect centres in 2H-MoS$_2$, in 2D and in nano-crystalline phase alike. In our previous work, we identified three paramagnetic defects in powdered 2H-MoS$_2$ nanocrystals synthesised through hydrothermal route, and we assigned them as molybdenum antisites, sulfur vacancies and oxygen vacancies of reduced molybdenum oxide\cite{khamrui2023study}. There are several theoretical studies in the literature addressing these intrinsic point defects in the monolayer, few layer vs. bulk counterparts \cite{Tsunetomo2021, han2016investigation, zheng2014tuning}. However, the role of external strain behind charge localisation at the same footing to the known experimental observations of nano-structured 2H-MoS$_2$ has not been elucidated. The structural strain may alter the singlet occupancy of electrons in the $d$-orbitals of Mo$^{4+}$($4d^2$) ions in pristine 2H-MoS$_2$ bulk nano-crystalline phase. Despite several experimental and theoretical modelling of searching charge-centres \cite{Cortes2018, dolui2013possible, HOUSSA2017, tan2020stability}, there is still lack of clarity on the origin, spin state and dynamics of intrinsic localised spin-centres in nano-crystalline 2H-MoS$_2$.  

In this work, we have reported low temperature EPR measurements on a series of hexagonal sulfur-deficit molybdenum disulfide (2H-MoS$_{2-x}$) nanocrystals to study the nature and evolution of various intrinsic paramagnetic defects. Also, we took aid of the first-principle calculations to understand the spin-states and found the favourable condition for spin-localisation around the defect centres, which has been further supplemented by experiments. Finally, with the help of spin-echo measurements, we have found differences in the spin relaxation behaviour of these defect centres. Furthermore, a particular set-up in the DFT exchange-correlation unveiled the role of nano-structured induced critical in-plane and out-of-plane strains on 2H-MoS$_{2-x}$, as the possible player behind the evolution in the spin-states of these defect centres. 

\section{Results and discussion}
\subsection{Material Synthesis and Characterization}
\begin{figure}[h!]
 \centering
 \includegraphics[width=\columnwidth]{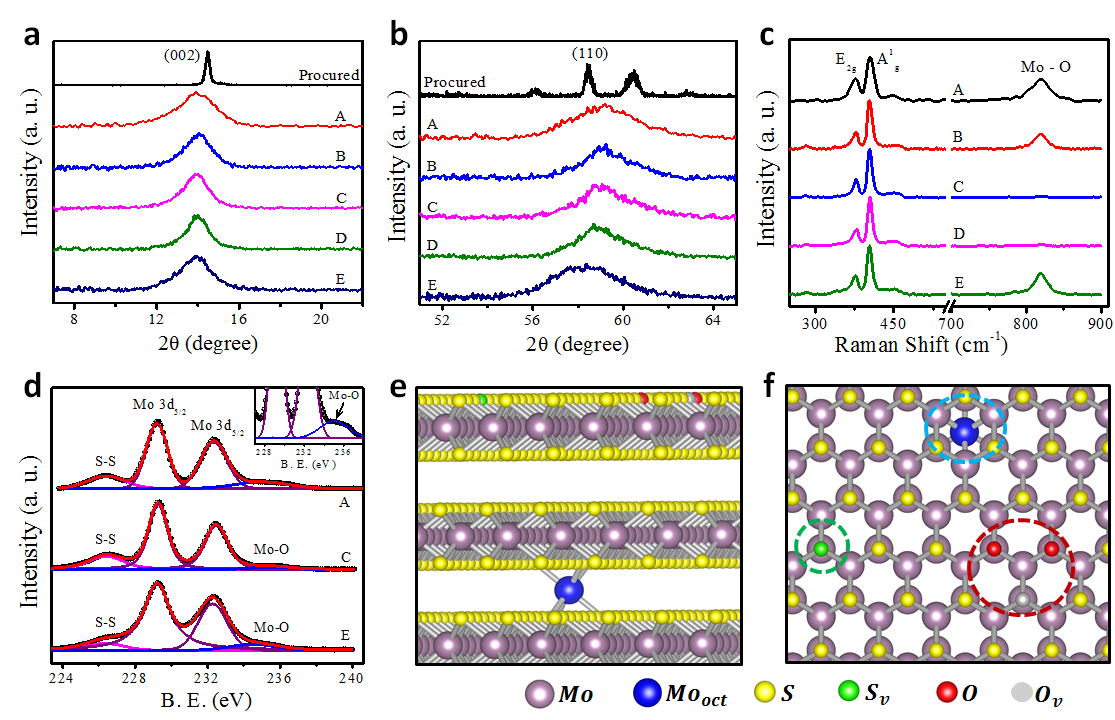}
 \caption{(a, b) XRD spectra of the as-synthesised MoS$_{2-x}$ nanocrystals and procured 2H-MoS$_2$ sample (standard) showing (002) and (110) peaks. (c) Raman spectra of all the as-synthesised samples. (d) Mo (3d) XPS spectra of sample A, C and E. Inset shows the zoomed in de-convoluted Mo(3d) plot of sample A showing the pronounce signature of Mo-O bonds. (e-f) A schematic of the multi-layered nano-crystalline 2H-MoS$_{2-x}$ samples has been shown with the side view and top view, respectively. All possible defect centres are encircled with dotted coloured circles.}
 \label{fig:1}
\end{figure}

MoS$_{2-x}$ nanocrystals were synthesised through hydrothermal method. In the synthesis, molybdenum trioxide (MoO$_3$) and thiourea (CH$_4$N$_2$S) were mixed at the molar ratios of 1:1, 1:2, 1:5, 1:8 and 1:10 to prepare five different batches of as-synthesised MoS$_{2-x}$ nanocrystals, namely A, B, C, D and E. Sample synthesis procedure has been described in detail in the Experimental section. XRD measurements (\textbf{Figure \ref{fig:1}a}, \textbf{b}) of these as-synthesised MoS$_{2-x}$ nanocrystals revealed hexagonal crystallographic phase formation with an out-of-plane tensile strain in comparison to the standard MoS$_2$ sample (procured from Sigma-Aldrich, India). This tensile strain was found to remain invariant ($\sim$ 3.2\%) among all the five as-synthesised samples. However, it decreased to 2.84\% for sample C1, which is the annealed form of sample C in Ar environment. On the other hand, a very small amount of in-plane strain ($\sim$ 1\%) was also found to be present in our samples. The crystallite sizes were found to range in between 3.94 nm to 7.27 nm, which refers to the bulk phase of the synthesised 2H-MoS$_{2-x}$ nanocrystals. However, these crystallite sizes remained very less (3 - 6 $nm$) in comparison to standard MoS$_2$ sample (33 $nm$). The values of lattice parameters, crystallite sizes and calculated strains are listed in \textbf{Table \ref{tbl:table1}}. A full range XRD spectra of the synthesised 2H-MoS$_{2-x}$ nanocrystals can be found in \textbf{Figure S1} in supplementary information to compare the several other crystallographic peaks with the standard MoS$_2$ sample.
 
Raman spectroscopic measurements (\textbf{Figure \ref{fig:1}c}) revealed in-plane ($E_{2g}$) and out-of-plane ($A^1_g$) Raman modes situated at 377.8 cm$^{-1}$ and 403.3 cm$^{-1}$, which supported the 2H-crystallographic phase formation of the MoS$_{2-x}$ nanocrystals. Apart from these two major Raman modes, there is another Raman peak appearing around 816 cm$^{-1}$, which arises from the Mo-O bonds \cite{vasilopoulou2012influence}. X-ray photoelectron spectroscopy (XPS) measurements further verified the hexagonal phase formation of the MoS$_{2-x}$ nanocrystals. \textbf{Figure \ref{fig:1}d} show the de-convoluted Mo(3d) core level XPS spectra of sample A, C and E indicating the presence of S-S, Mo-S and Mo-O bonds. The Mo-atom in the Mo-O bonds were found to remain in the $\delta ^+$-charge state (4 $\textless$ $\delta$ $\textless$ 6), which indicates the presence of unreacted molybdenum oxysulfide (MoS$_y$O$_z$) as impurity in the sample \cite{vasilopoulou2012influence}. The other samples also showed similar atomic bonds, but with varying fraction as investigated in detail in our previous work \cite{khamrui2023study}. Also, the sulfur to molybdenum atomic ratio was found to lie in between 1.59 to 1.95, which indicates the sulfur-deficit phase formation of MoS$_2$. TEM measurements were carried out to probe the atomic structure of our synthesised MoS$_2$ nanocrystals microscopically. \textbf{Figure S2} shows the TEM images of sample A, C, E and C1. It can be seen that there are many stacked atomic planes appeared as parallel lines or crossed each other. Also, the parallel planes are not continuous for a long distance as broken regions and voids are present, which are marked by red dotted boxes.

Based on the above discussed spectroscopic data, we can say that the lattice of our synthesised nanocrystals are mostly 2H-MoS$_{2-x}$ type. Though there are the presence of MoS$_y$O$_z$ as impurity, it does not appear as separate island. Instead, it stays distributed within the 2H-MoS$_{2-x}$ matrix and interconnected among themselves. \textbf{Figure \ref{fig:1}e} and \textbf{f} shows the schematics to demonstrate the lattice formed in our samples. Considering sulfur-deficit to sulfur-rich growth conditions, there are three possible major defect centres, namely molybdenum interstial ($Mo_{int}$) residing in the van der Waals spacing, sulfur vacancy ($S_v$) of 2H-MoS$_{2-x}$, and oxygen vacancy ($O_v$) of MoS$_y$O$_z$.

\begin{table}
  \caption{List of lattice constants ($a,c$), atomic ratio (S:Mo) and crystallized size (D) of the as-synthesised and annealed 2H-MoS$_{2-x}$ nanocrystals. Lattice strain has been calculated with respect to the procured 2H-MoS$_2$ (considered as standard sample).}
  \label{tbl:table1}
  \begin{tabular}{cccccc}
    \hline
    Sample & S/Mo & a ($\AA$)&c ($\AA$)& Crystal Size, D & Strain\\ 
    Name   & ratio&         &          & (nm)            & [002] \\ 
           &      &         &          &                 & (\%)  \\\hline\hline
    A      & 1.62 &  3.12   &  12.68   &3.94             & 3.17  \\
    B      & 1.59 &  3.12   &  12.67   &5.11             & 3.09  \\
    C      & 1.74 &  3.13   &  12.69   &5.75             & 3.25  \\
    D      & 1.88 &  3.13   &  12.69   &7.27             & 3.25  \\ 
    E      & 1.95 &  3.16   &  12.69   &4.93             & 3.25  \\ 
    C1     & 1.66 &  3.13   &  12.64   &6.91             & 2.84  \\
    Procured& -   &  3.16   &  12.29   &33.00            & 0     \\ \hline
  \end{tabular}
\end{table}

\subsection{Paramagnetic Defects in 2H-MoS$_2$}
\begin{figure}[h!]
 \centering
 \includegraphics[width=\columnwidth]{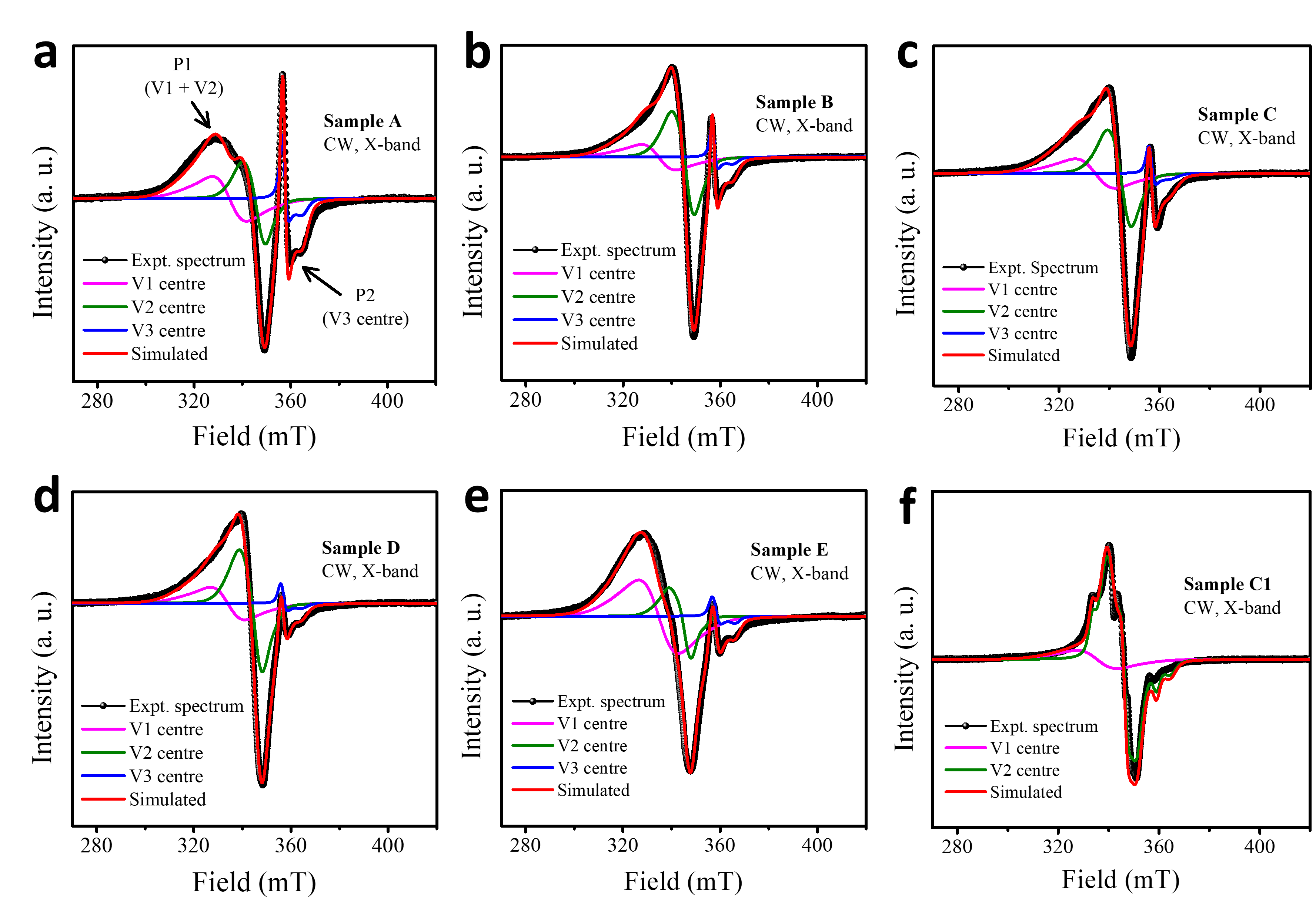}
 \caption{X-band CW-EPR spectra recorded at 12K. Black curves represent the experimentally acquired data. Each experimental spectrum has been matched with the simulated one (red curves) by considering three spin components, which are defined as V1 centre (magenta), V2 centre (olive) and V3 centre (blue).}
 \label{fig:2}
\end{figure}

We have conducted low temperature CW-EPR measurements on the 2H-MoS$_{2-x}$ nanocrystals to understand the nature and evolution of the above mentioned intrinsic defects with varying growth condition. \textbf{Figure \ref{fig:2}(a-e)} shows the CW-EPR spectra of all samples along with the simulated one which adequately fit the experimental data. EPR spectra simulations were performed using EasySpin spectrum simulation toolbox in Matlab environment \cite{stoll2006easyspin}. In the simulation, we considered three spin centres, namely V1 (S = $\frac{3}{2}$), V2 (S = $\frac{3}{2}$) and V3 (S = $\frac{1}{2}$). The rationale for choosing the above mentioned spin states has been discussed in the sections of Rabi nutation measurement and DFT spin-states calculation.

The variation in relative weight of the three spin centres has been listed in \textbf{Table \ref{tbl:table2}}. It is seen that the relative weight of V1 centre ranges in between 14-18\% for sample A, B, C and D, while it increased almost two times for sample E (43\%). At the same time, the relative weight of V2 centre ranges in between 51-60\% for sample B, C and D, while it falls down by half for sample A and E. On the other hand, the relative weight of V3 centre decreased almost monotonically from 52\% to 23\% for sample A to E. The sulfur-deficit growth condition leads to the maximization of V3 centre in relative weight, while extreme sulfur-rich growth condition leads to the  maximization of V1 centre. After annealing, V2 centre becomes majority, with a weak presence of V1 centre and disappearance of V3 centre (sample C1 in \textbf{Table \ref{tbl:table2}}). The Lande g-factors of the 3 centres (V1: g$_x$ = g$_y$ = g$_z$ = 2.062, V2: g$_x$ = 2.035, g$_y$ = 2.000, g$_z$ = 1.984 and V3: g$_x$ = g$_y$ = 1.934, g$_z$ = 1.892) revealed an isotropic symmetry for the V1 centre, rhombohedral symmetry for the V2 centre and axial symmetry for the V3 centre. Also, the zero field splitting (ZFS) parameters were taken equal in the simulation of all the as-synthesised samples ($\mathscr{D}$ = 320 MHz, $\mathscr{E}$ = 30 MHz for V1; $\mathscr{D}$ = 80 MHz, $\mathscr{E}$ = 20 MHz for V2). However, the ZFS of sample C1 adapted almost a three times higher value as compared to sample C to match the fine structures of the EPR lineshape ($\mathscr{D}$ = 180 MHz, $\mathscr{E}$ = 58 MHz for V2). This increment in ZFS value may be due to the change in the local structure of MoS$_2$ around V2 centre in the enhanced crystalline phase. More details of the EasySpin simulation can be found in \textbf{Table S1}, \textbf{S2} and \textbf{S3} in SI. Based on the symmetry of g-factor and literature reports \cite{HOUSSA2017,wagner1979epr,cai2015vacancy}, the V1 and V2 centres can be ascribed as molybdenum interstitial and sulfur vacancy of 2H-MoS$_{2-x}$ phase, while the V3 centre can be assigned as oxygen vacancy of MoS$_y$O$_z$ phase.
 
\begin{table}
  \caption{List of the relative weight for V1, V2 and V3 spin centres as calculated from CW-EPR spectra. Also, the values of spin-lattice relaxation times (round-off values) are listed for P1 and P2 as measured by inversion recovery technique at 12K.}
  \label{tbl:table2}
  \resizebox{\columnwidth}{!}{
  \begin{tabular}{ccccccccc}
    \hline
Sample & V1 centre  & V2 centre  & P1   & P2     &T$_1$ ($\mu s$)&T$_1$ ($\mu s$)&T$_1$ ($\mu s$)&T$_1$ ($\mu s$)\\ 
Name   & (\% weight)& (\% weight)& (V1+V2)   & V3 centre   & P1        & P1        & P2        & P2       \\ 
       &            &            &(\% weight)& (\% weight) & X-band        & W-band        & X-band     &  W-band \\ \hline\hline
A      &  18        &  30        &  48       & 52          & 228         &  500          & 382         &   1005     \\
B      &  14        &  51        &  65       & 35          & 163         &  550          & 381         &   1130     \\
C      &  17        &  51        &  68       & 32          & 180         &  400          & 352         &   1000     \\
D      &  18        &  60        &  78       & 22          & 129         &  360          & 441         &   1010     \\ 
E      &  43        &  34        &  77       & 23          &  62         &  650          & 605         &   1020     \\ 
C1     &   8        &  92        & 100       & -           & 299         &  510          &   -           &    -       \\ \hline
  \end{tabular}
}
\end{table}

\subsection{Spin Relaxation Measurements of the Paramagnetic Defects}
With the different local symmetry, the defect centres may posses distinct spin relaxation behaviour. In our previous work, we found a significant difference in spin-spin relaxation time-scale among the spin centres \cite{khamrui2023study}. To further probe into their relaxation behaviour and dynamics, we have measured spin-lattice relaxation (T$_1$) times at the positions of P1 and P2 at 12 K for all samples. We employed inversion recovery method with 2 step phase cycling to measure T$_1$ of both P1 and P2 in the X- and W-band frequencies (pulse sequence is depicted in \textbf{Figure S5}). A bi-exponential growth curve, consisting of a short and a long time scale, was used to fit the inversion recovery data (\textbf{Figure S7}). The longest time scale can be identified as the spin-lattice relaxation time, whereas the short time scale comes from the fast relaxation processes such as spin diffusion, spectral diffusion etc. Values of T$_1$ for both P1 and P2 are listed in \textbf{Table \ref{tbl:table2}}. It can be seen that in general, T$_1$ value of P2 (corresponds to MoS$_y$O$_z$ phase) is much longer than T$_1$ of peak P1 (corresponds to 2H-MoS$_{2-x}$ phase) in the X-band frequency. 

\begin{figure}[h!]
 \centering
 \includegraphics[width=\columnwidth]{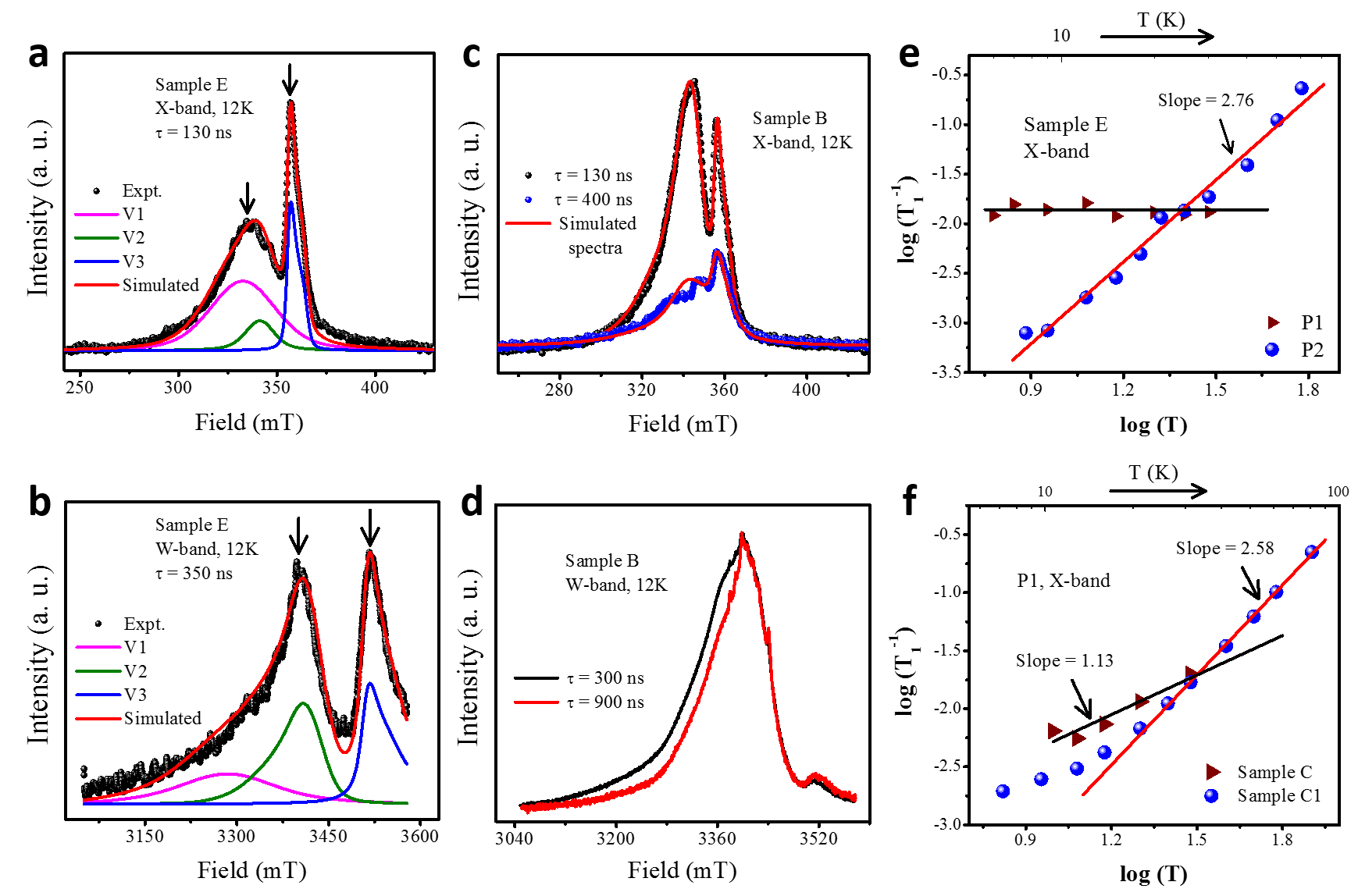}
 \caption{(a, b) Echo detected field sweep spectrum of sample E measured at 12K in the X- and W-band frequencies. The vertical down arrows mark the positions where T$_1$-relaxations were recorded. (c, d) Echo detected field sweep spectrum of sample B measured at the X- and W-band frequencies at 12K. (e) Plot of spin-lattice relaxation rate with temperature for sample E measured at the position of P1 and P2 in the X-band frequency. (f) The same for sample C and C1 measured at the position of P1.}
 \label{fig:4}
\end{figure}

In the W-band, Larmor frequency increases $\sim$ 10 times, so as the mutual separation in field position of two signals originating from two spin centres. It is evident from comparison of spectra for the same samples at same temperature at two frequencies as shown in \textbf{Figure \ref{fig:4}} (compare panels \ref{fig:4}a with \ref{fig:4}b and \ref{fig:4}c with \ref{fig:4}d). As a result, the T$_1$ measurements in the W-band at a given position is less influenced by other spin centres as compared to the X-band. In the enhanced spectral resolution of the W-band, the T$_1$-values of P2 (V3 centre) was found to remain almost constant for all the as-synthesised samples ($\sim$ 1 ms). While the T$_1$ value at peak P1 of sample E becomes 650 $\mu s$, which is comparable (actually highest) with the T$_1$ value of other samples, as opposed to the X-band results where T$_1$ of peak P1, sample E is lower by an order as compared to other samples. To asses this issue, we looked at the echo detected field sweep spectra of sample E in the X- and W-band frequency as shown in \textbf{Figure \ref{eq:4}}a and b. The vertical down arrows show the positions where T$_1$-relaxations were recorded. It can be seen that the V1 centre dominates over V2 for the T$_1$ measurement of P1 in X-band frequency, while the V2 centre dominates peak P1 in the W-band frequency at the position of T$_1$ measurement. So, there is an indication that the inherent T$_1$-relaxation time of V1 centre is much lower than that of the V2 centre. To further verify this argument, we measured T$_1$-relaxation times for the annealed sample C1 in the X- and W-band frequencies, where the V2 centre is present in majority (V1: 8\%, V2: 92\%) and V3 centre is absent. From Table \ref{tbl:table2}, we can see that the T$_1$-value of P1 increases almost 1.5 times after annealing in both frequency bands, which clearly indicates to a slower relaxation process in V2 as compared to V1. Similarly, a careful component-wise analysis of all the samples reveal the following general trend: the V3 centre (peak P2) has longest T$_1$, whereas V1 centre (left side component of peak P1) has the shortest T$_1$ values for any sample ($T_1^{V_3} \:\textgreater \: T_1^{V_2} \:\textgreater \: T_1^{V_1}$).

To understand the spin-spin relaxation of the defect centres, we measured inter-pulse time ($\tau$) dependent field sweep spectra of sample B with a Hahn echo sequence \cite{hahn1950spin} ($\pi$/2 - $\tau$ - $\pi$ - Echo) in the X- and W-band frequencies as shown in \textbf{Figure \ref{fig:4}}c and d. It can be seen that the intensity loss at the position of V1 centre is greater than that of V2 and V3 centres, which is evident both in the X- and W-band measurements. To get an estimate of the intensity loss, we simulated the echo detected field sweep spectra for the two $\tau$-values at X-band. We found that the intensity ratio between V1 and V2 decreased from 0.45 ($\tau$ = 130 ns) to 0.32 ($\tau$ = 400 ns), which indicates that the echo decay is fastest in V1 with shortest spin phase memory time $T_M$. As echo decay is considered as a direct measure of spin-spin interaction in magnetic resonance, it may be concluded that this interaction is highest among the V1 defects and lowest among the V3 defects. So, the phase memory time of the defect centres can be arranged in the following order : $T_M^{V_3} \:\textgreater \: T_M^{V_2} \:\textgreater \: T_M^{V_1}$.

To get further insight of spin-lattice relaxation dynamics of the defect centres, we performed temperature dependent T$_1$-relaxation measurements at the position of P1 and P2 for sample A, C, E and P1 center of C1 in the X-band frequency. Temperature dependent inversion recovery data can be found in \textbf{Figure S8}. Here, \textbf{Figure \ref{fig:4}}e shows log(T$_1^{-1}$) vs log(T) plot for sample E. It can be seen that relaxation rate increases linearly with an exponent of 2.76 for P2 (V3 centre), whereas it is almost independent of temperature for P1 (major contribution from V1 centre). On the other hand, post synthesis annealing has been found to increase the exponent value for P1 from $\sim$ 1 (sample C, V1: 17\%, V2: 51\%) to $\sim$ 2.6 (sample C1, V2: 92\%) as shown in \textbf{Figure \ref{fig:4}}f. At the same time, temperature dependent T$_1$ measurements of sample A (\textbf{Figure S9}) revealed the exponents values as 1.1 and 2.1 for P1 and P2, respectively. 

There are three major routes of spin-lattice relaxation by which an unpaired electron spin and lattice equilibrate, namely direct, Raman and Orbach process. \cite{lumata2013electron, rao2012spin}. In direct process, T$_1$-relaxation rate increases linearly or remains almost independent of temperature. In case of Raman relaxation, T$_1$-relaxation rate is proportional to T$^2$ well above the Debye temperature (T$_D$), whereas it varies with T$^x$ ($3 \textless x \textless 9$) below T$_D$. On the contrary, the T$_1$-relaxation rate in Orbach process varies as 1/T. Based on our observations, it can be said that the V2 and V3 centres follow Raman-like relaxation mechanism, whereas the  V1 centre most likely follows a direct process for spin-lattice relaxation. So, it can be concluded that the vacancy defects (V2 and V3) are relatively slow relaxing spin centres following Raman-like relaxation mechanism as compared to the interstitial defects (V1), which follow direct spin-lattice relaxation mechanism at low temperatures ($\textless$ 85K).

\subsection{Hyperfine Measurement of the Defect Centres}
\begin{figure}[h]
 \centering
 \includegraphics[width=\columnwidth]{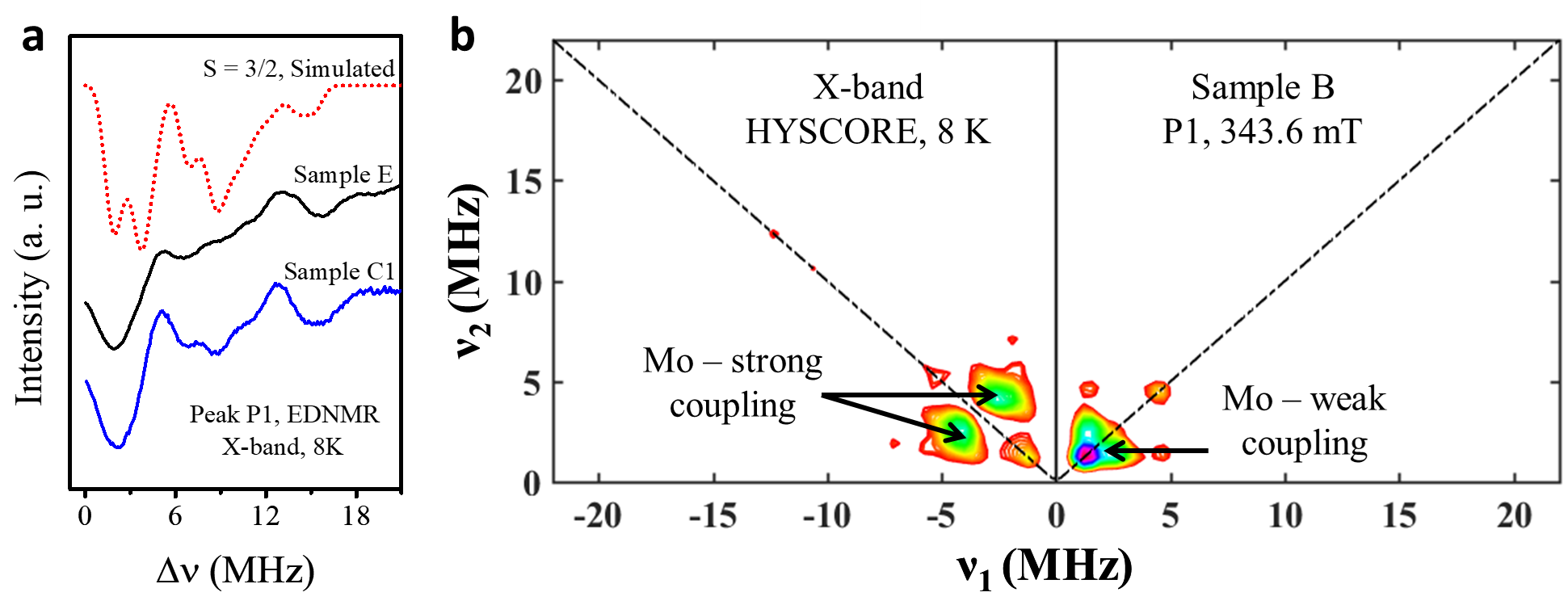}
 \caption{(a) X-band EDNMR spectra of sample E and C1 measured at the position of P1 at 8K. Red dotted curve represents the possible nuclear transitions simulated by the 'Salt' package of the EasySpin simulation toolbox. (b) X-band HYSCORE spectrum of sample B measured at the position of P1 (343.6 mT) and at the temperature of 8K.}
 \label{fig:3}
\end{figure}

To understand the nuclear environment of the spin centres, we employed X-band electron-electron double resonance (ELDOR) detected NMR technique. The pulse sequence of this measurement has been depicted in \textbf{Figure S3}a. Here, \textbf{Figure \ref{fig:3}}a shows the ELDOR-detected-NMR (EDNMR) spectra of sample E (highest content of V1) and C1 (mostly V2) measured at the position of peak P1. Multiple nuclear transitions are found to appear in the EDNMR spectra, which can be simulated by considering the $^{95}$Mo [A$_{hyperfine}$ = (5.0 5.0 8.0) MHz, Q$_{quadrapole}$ = (-0.04 -0.04 0.08) MHz] and $^{1}$H [A$_{hyperfine}$ = 0.55 MHz] nuclei coupled with a spin-$\frac{3}{2}$ centre. If we consider spin-$\frac{1}{2}$ centre, the broad feature in the range 6 - 12 MHz of EDNMR spectra is not reproduced, whereas considering spin-$1$, the low frequency EDNMR doublet (0 - 6 MHz) was absent.  More details of the simulation and EDNMR spectra of other samples can be found in \textbf{Figure S3(b, c)}. So, these results indicate that peak P1 (cumulative contribution of V1 and V2 centres) possess spin-$\frac{3}{2}$ state.

Further understanding of the nuclear environment comes from the Hyperfine Sublevel Correlation Spectroscopy (HYSCORE) technique (pulse sequence is depicted in \textbf{Figure S4a}), where correlations of hyperfine interactions are better resolved. \textbf{Figure \ref{fig:3}}b shows the HYSCORE spectrum of sample B measured at the position of P1 in the X-band frequency. It can be seen that there is a pair of cross peaks in second quadrant centred around (-3.31, +3.31) MHz, which indicates strong hyperfine coupling of $^{95}$Mo with the electron spin for peak-P1. We also conducted HYSCORE measurements for sample A (at the position of P2) and sample C1 (at the position of P1), and we identified the similar strong hyperfine coupling of $^{95}$Mo (see \textbf{Figure S4b} and \textbf{c}). Though $^{97}$Mo has Larmor frequency close to $^{95}$Mo, it can be neglected as it has larger quadrupole moment which could led to extensively broadened hyperfine ridges \cite{goldberg2009oxidation}. Apart from the $^{95}$Mo strong coupling, there is one peak centred at (+1.20, +1.20) MHz on the diagonal of first quadrant, which may arise from weak hyperfine coupling of distant Mo nucleus. As g-factor of the V1 and V2 centres are close to each other, it is difficult to probe their nuclear environment separately in the X-band frequency. In our samples, peak P1 contains hyperfine information of both V1 and V2, while peak P2 gives hyperfine information of mostly the V3 centre. So, the appearance of strong hyperfine coupling due to $^{95}$Mo nuclei further indicates that all the three localised spins are Mo-centric.

\subsection{Rabi Oscillation Measurement}
\begin{figure}[h!]
 \centering
 \includegraphics[width=\columnwidth]{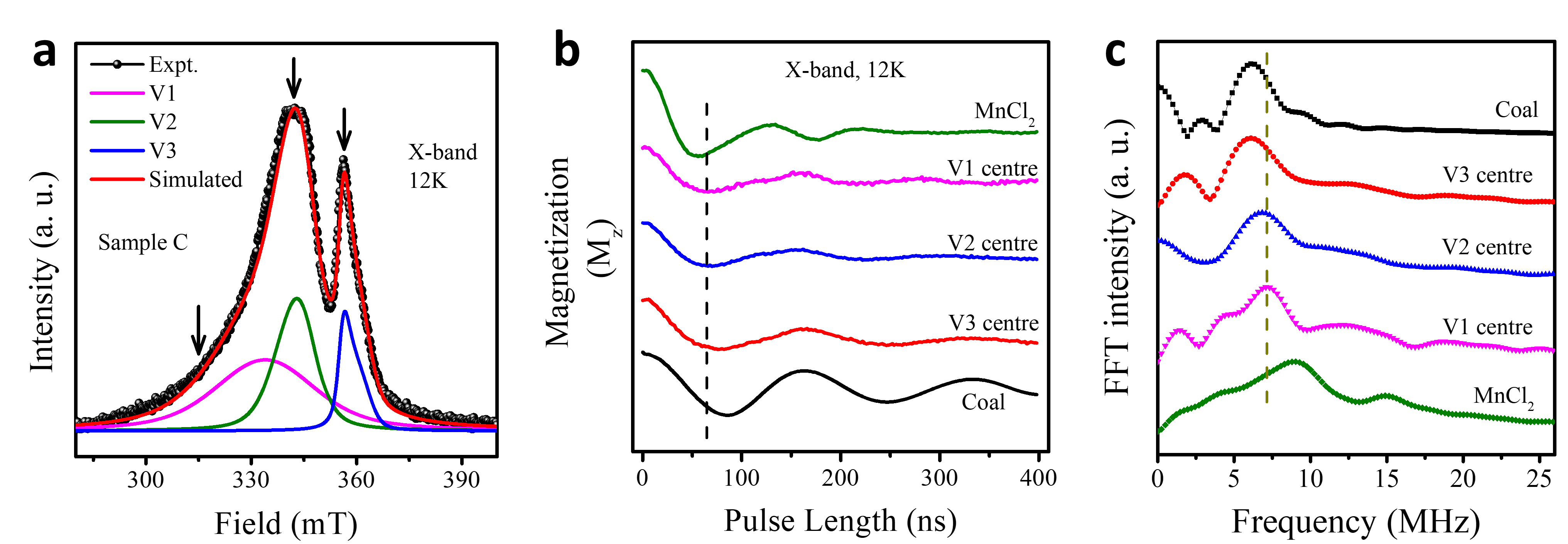}
 \caption{(a) X-band echo detected field sweep spectrum of sample C measured at 12K. The vertical down arrows show the positions where Rabi oscillations were recorded. (b) Rabi oscillation plot of the coal sample, MoS$_{2-x}$ nanocrystals (sample C) and MnCl$_2$ sample measured at 12 K with the nutation pulse power of 2dB in the X-band frequency. Vertical dotted line shows the average position of first minima of each nutation curve measured at the marked positions of sample C. (c) Fourier transform of the Rabi oscillation data. The dashed vertical line marks the peak position for the V1 and V2 centre.}
 \label{fig:5}
\end{figure}

While looking at the W-band field sweep spectra, it can be seen that the entire EPR lineshape has a spread of $\sim$ 550 mT, which is much higher than the expected magnetic field span for three spin-$\frac{1}{2}$ species appearing side-by-side in the spectrum (EPR linewidth for a typical spin-$\frac{1}{2}$ signal is $\sim$ 5 mT). It further indicates that the localised spin centres of our synthesised 2H-MoS$_{2-x}$ may be in high spin states. To determine the possible spin states of the defect centres, we performed Rabi nutation measurements at the position of V1, V2 and V3 centres for sample C, which are then compared with the nutation of standard spin-$\frac{1}{2}$ (coal sample from Bruker Biospin GmbH) and spin-$\frac{5}{2}$ (0.8  mM frozen solution of MnCl$_2$) samples as shown in \textbf{Figure \ref{fig:5}b}. Field swept spectra and further details of the standard samples can be found in \textbf{Figure S10} and subsequent text. In this technique, a three pulse sequence was used (depicted in \textbf{Figure S11}), where the length of nutation pulse was incremented by 2 ns step and the oscillation of magnetisation was recorded through detection pulse as a function of this increasing nutation pulse length. For spin S, the nutation frequency of the transition between m$_s$ to (m$_s$ $\pm$ 1) can be written as - 
\begin{equation}
\omega_{m_s, m_s \pm 1} = \omega_1 \sqrt{S(S + 1) - m_s(m_s \pm 1)}
\end{equation}
where $\omega_1$ = g$\mu_B$B$_1$/$\hbar$ and B$_1$ is magnetic field of the microwave. 
From equation (1), it can be derived that the ratio of nutation frequency for spin-$\frac{5}{2}$, $\frac{3}{2}$ and $\frac{1}{2}$ will be 3 : 2 : 1, respectively for the central transition (m$_s$ = -$\frac{1}{2}$ to $\frac{1}{2}$). 

It can be seen that the first minima of all the defect centres in MoS$_{2 - x}$ fall in between spin-$\frac{5}{2}$ MnCl$_2$ and spin-$\frac{1}{2}$ coal. To calculate the accurate nutation frequencies, we did Fourier transform of the Rabi nutation data as shown in \textbf{Figure \ref{fig:5}c}. It is observed that nutation frequency of the V1 and V2 centres fall nearly at the middle of MnCl$_2$ (S=$\frac{5}{2}$) and coal sample (spin S=$\frac{1}{2}$), while the nutation frequency of the V3 centre exactly matches with that of coal. So, most likely V1 and V2 defect centres posses either a spin-$1$ or a spin-$\frac{3}{2}$ state. In connection to the EDNMR measurements and related simulations described in the previous section, it will be logical to conclude that the V1 and V2 centres posses spin-$\frac{3}{2}$ state, while the spin multiplicity of V3 centre is $\frac{1}{2}$. In our measurements, the ratio of the nutation frequencies for standard spin-$\frac{1}{2}$ and $\frac{5}{2}$ samples is 1.48, as opposed to the theoretical value of 3. It is due to the overlap of the allowed EPR transitions coming from other m$_s$-manifolds and forbidden transitions in the X-band frequency. 

\subsection{Analysis of Defect Centres from DFT Calculations}
\label{sec:dft1}
The present experimental observations on the origin and dynamics of the localized charge centres in the nano-structured 2H-MoS$_{2-x}$ are also modelled through the density functional theory based defect sites engineering to better understand their origin. In the DFT modelling we took into account few prime point defects, i.e. sulfur vacancy (S$_v$), Mo-interstitial at the tetrahedral (Mo$_{tet}$) and octahedral sites (Mo$_{oct}$) within the van der Waals spacing, and oxy-sulfide formation at $S_v$-sites ($V_{SO}$) with or without external strain (for unstrained cases schematic models are shown in \textbf{Figure S12} in SI). The single and multi S$_v$ formation energies are extensively studied in the earlier literature at different DFT exchange-correlation set-up on the pristine bulk phase or mono-layers of 2H-MoS$_2$ without external strain reference to the volume from present first-principles calculations \cite{komsa2015native,zhou2013intrinsic,Zhou2013}. We have calculated electronic- and spin-properties of these different point defects in the nano-crystalline 2H-MoS$_2$ mimicking the nano-structured induced strain in the as synthesized 2H-MoS$_{2-x}$ samples from present experimental data (see DFT Methodology section for more details). 

We have used Hubbard onsite-$U$ correction and vdW-dispersion energy (at D3-level correction) in the choice of the exchange-correlation functional (see also \textbf{Figure S13} and \textbf{Figure S14} in SI). In the proposition of a exchange-correlation functional through PBE-D3+$U$ method with effective U = 4.0 eV for both Mo(4d) and S(3p) or O(2p) orbitals are opted based on the calculated electronic band structures and thermodynamics of defect formations energies in the pristine phase (See \textbf{Table S5} in SI and also, DFT Methodology Section). Despite the marginal improvement (5-10\%) of the band-gap of 2H-MoS$_{2}$ which falls within the hybrid DFT calculations error bars of quasi-2D solids\cite{dasJCTC2019,Borlido2020npj,Yang2023npj}, the choice remain reasonable as it does not deteriorate the earlier DFT predicted formation of the single sulfur vacancy formation energies, $E_{form}$ = 2.16-2.18 eV per vacancy in comparison to the earlier DFT data 2.0-3.5 eV, but at the modest computational cost (see \textbf{Table S6} in SI)\cite{dolui2013possible,Cortes2018,tan2020stability}. 

\begin{figure}
 \centering
 \includegraphics[width=\columnwidth]{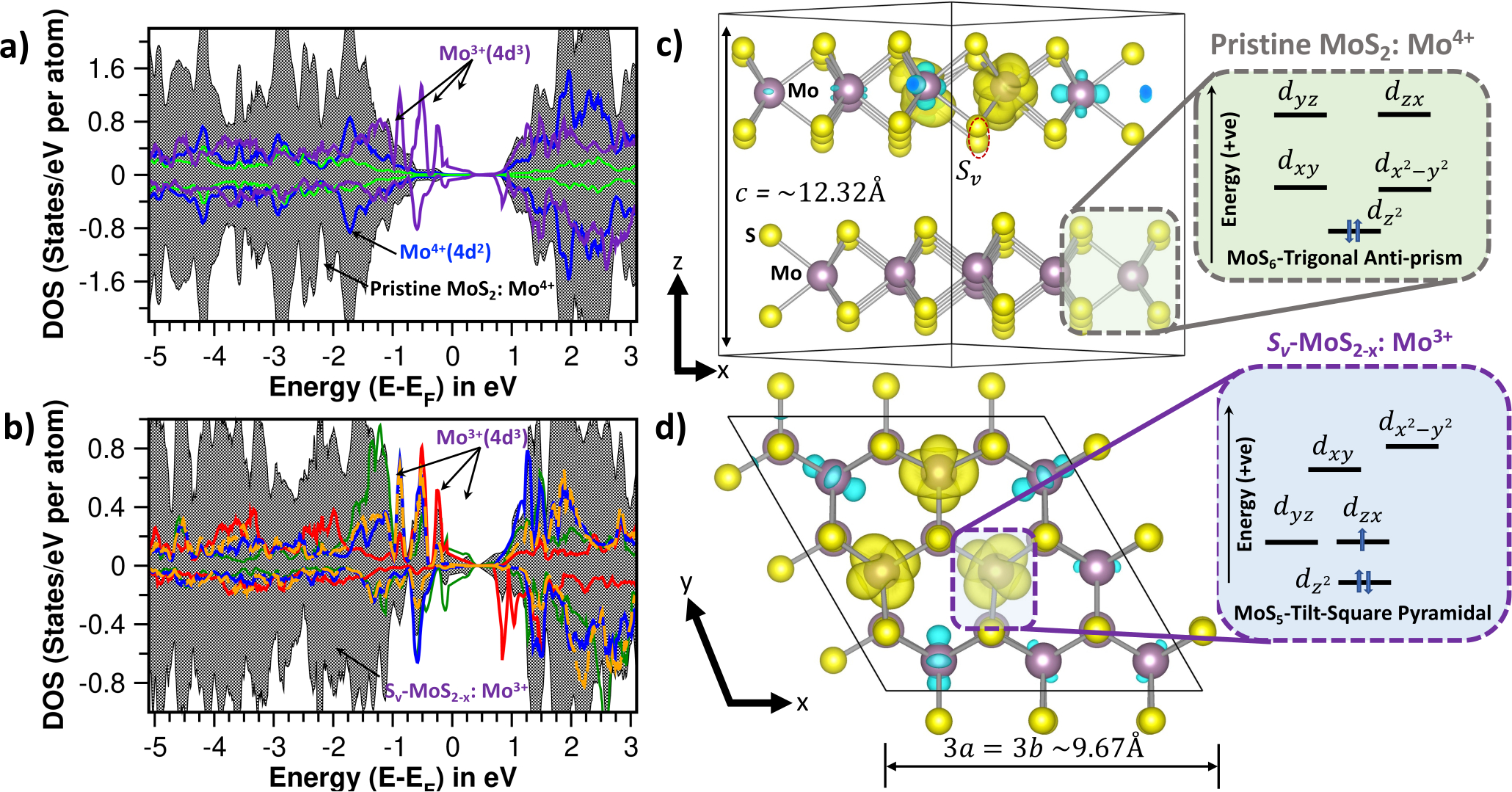}
 \caption{In the left panel a) majority and minority total density of states (DOS) of the pristine 2H-MoS$_2$, d-orbital of Mo$^{4+}$(4d$^2$) and S(3p) projected DOS (pDOS) are shown with black shaded areas, blue and light green solids lines, respectively while only pDOS of the Mo$^{3+}$(4d$^3$) is shown with the solid thick indigo color lines in this panel. In the panel b) similarly, only the Mo$^{3+}$(4d$^3$) orbital projected pDOS is shown, where total DOS of the 2H-MoS$_{2-x}$ is plotted and marked with black shaded area, and four other possible non-degenerate Mo-orbitals of Mo$^{3+}$(4d$^3$) are shown with color lines: Mo-4d$_{yz}$/d$_{xz}$ in red, Mo-4d$_{z^2}$ in deep green and Mo-4d$_{xy}$ in orange dashed lines, and Mo-4d$_{x^2-y^2}$ in blue solid lines. Corresponding spin-density plots for the Mo$^{4+}$(4d$^2$) and Mo$^{3+}$(4d$^3$) with their formal charge center are shown in the panels c) and d) with an IOS level 0.01 e$^-$ per $\AA$ $^3$. Possible 4d-orbital occupancies schema are zoom out in the right side boxes with up and down arrows for majority and minority spins of Mo$^{3+}$(4d$^3$) electrons, respectively.}
 \label{fig:Vs}
\end{figure}
The tilted square-pyramidal structure of MoS$_5$ motifs around the $S_v$ vacancy model deviated from an ideal anti-prism MoS$_6$ resulted three mono electron-polarons formed at the 3 nearest neighboring in-plane triangular Mo sub-lattices with an unpaired electron per site (See \textbf{Figure \ref{fig:Vs}}). The critical strain (2-6\% in-plane and 4\% out-of-plane tensile strain) generated from the nano-structuring are expected to be very essential and play a vital role for such Mo$^{3+}$(4d$^3$) spin-centres generation in the bulk phase of 2H-MoS$_{2-x}$ samples. These polaronic features remain silent in the unstrained pristine bulk materials as obtained from the current first-principles calculated volume of 2H-MoS$_2$ (Similar figure on unstrained pristine 2H-MoS$_2$ from QE.6.8 calculations is shown in \textbf{Figure S15} in SI), and consistent with the earlier predictions\cite{dolui2013possible,tan2020stability}. Similarly, the choice of the uni-axial tensile or compressive strain (2-4\%) on the $S_v$ included model or even 2-6\% in-plane and 4\% out-of-plane tensile strain oxy-sulfide ($V_{SO}$) terminated model neither capable to get the Mo$^{3+}$(4d$^3$) charge centres in 2H-MoS$_{2-x}$ (see SI \textbf{Figure S16}). 

In the present case of $S_v$ included nano-structured 2H-MoS$_{2-x}$ sample, total DOS and pDOS are shown in the left panels, \textbf{Figure \ref{fig:Vs}a)} and \textbf{b)} while, corresponding spin-density plots (yellow/blue bubbles on Mo,S-sites) for the formal Mo$^{4+}$(4d$^2$) and induced Mo$^{3+}$(4d$^3$) ions are shown in the panels \textbf{Figure \ref{fig:Vs}c)} and \textbf{d)} with side and top views, respectively. A schematic sketch on their possible Mo$^{3+}$(4d$^3$)-level occupancies is zoomed out on the shaded right side panels of this figure. Indeed, the $S_v$ centres related MoS$_5$ motifs from trigonal anti-prism configuration from ideal MoS$_6$ motifs eventually led to the occurrence of three single electron-polarons at the 3 nearest neighbouring Mo-sites resulted as doublet spin-state. Thus, the $S_v$ centre rolls out as 3 spin-$\frac{1}{2}$ centres interacting with a direct magentic-exchange coupling resulting pseudo Mo$^{3+}$ spin-centres. The net spin-magnetic moment 0.99 $\mu_B$ per Mo$^{3+}$(4d$^3$) sites is due to an unpaired electron per Mo$^{3+}$ site and spin-density plots clearly show such signature. From the current analysis, it looks like the possible source of the V2 centre as in the synthesized sample 2H-MoS$_{2-x}$, observed through the current exp. data analysis. 

\begin{figure}
 \centering
 \includegraphics[width=\columnwidth]{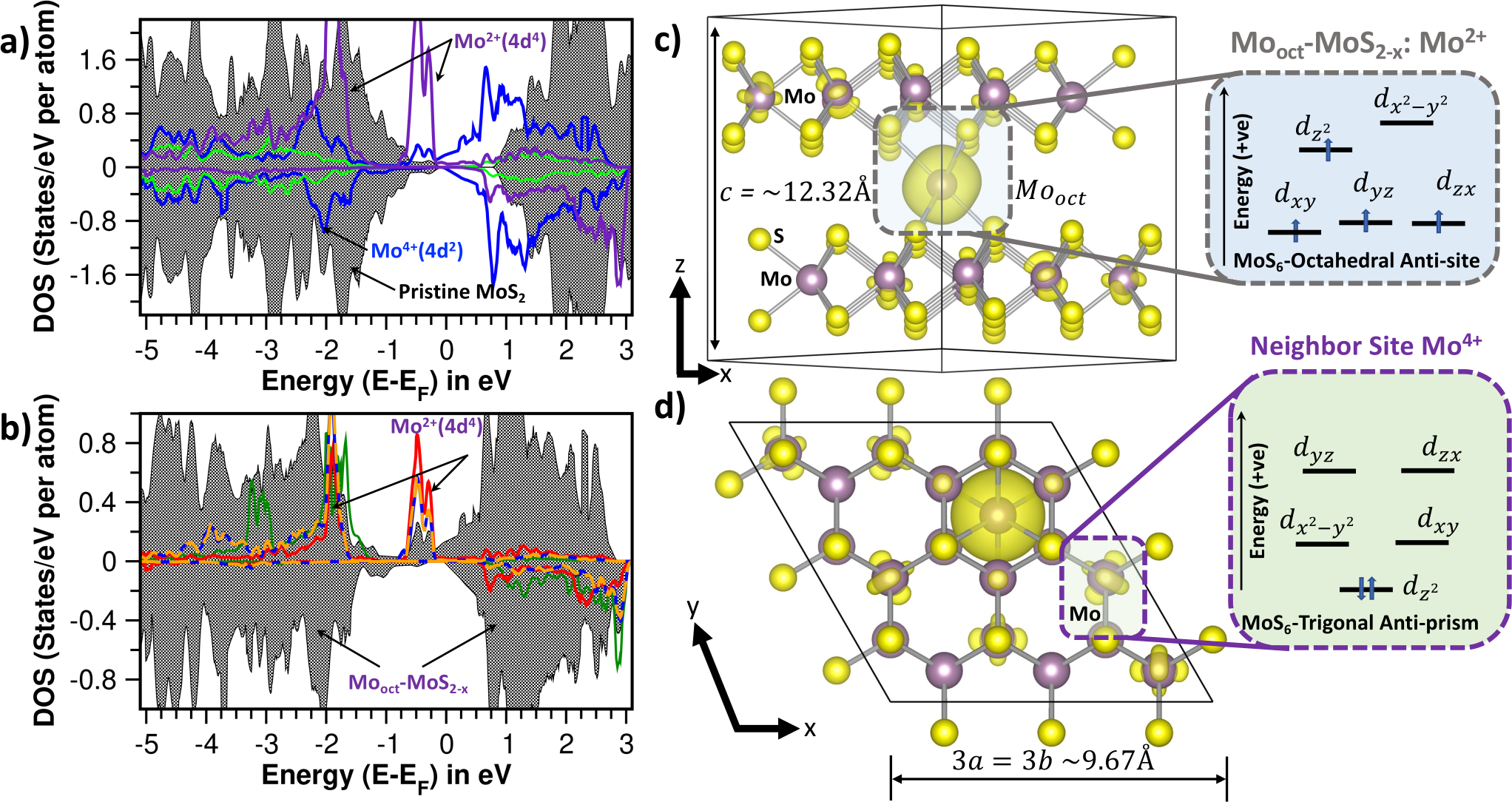}
 \caption{Single Mo interstitial at octahedral sites (Mo$_{oct}$) in the nano-crystalline bulk phase of 2H-MoS$_2$. In the left panel a) majority and minority total DOS of the pristine 2H-MoS$_2$, Mo$^{4+}$(4d$^2$) and S(3p) projected DOS are shown with black shaded areas, blue solid lines, light green solids lines, respectively, while the total 4d-orbital of the Mo$_{oct}$-site is shown with the solid thick indigo color solid line. The Mo$^{2+}$(4d$^4$) projected DOS is shown with its orbital decomposition in the panel b) where the total DOS of the S-deficit 2H-MoS$_{2-x}$ is presented with black shaded area, and other possible non-degenerate 4d-orbitals of Mo$^{2+}$(4d$^4$) are shown with color lines: Mo-d$_{yz}$/d$_{xz}$ in red, Mo-d$_{z^2}$ in deep green and Mo-d$_{xy}$/d$_{x^2-y^2}$ in orange/blue dashed or solid lines. Corresponding spin-density plots Mo-site with formal charge center Mo$^{2+}$(4d$^4$) and NN Mo$^{4+}$(4d$^2$) site is shown in the panels c) and d) with side and top views, respectively (3D IOS level 0.01 e$^-$/$\AA$ $^3$). Possible 4d-level occupancies schema are zoom out in the right side boxes with up and down arrows for majority and minority spins of Mo$^{2+}$(4d$^4$) electrons, respectively.} 
 \label{fig:MoOct}
\end{figure}
Similarly, possible other S-deficit 2H-MoS$_{2-x}$ models with additional Mo-interstitial is viable in the host 2H-MoS$_2$ material (See SI \textbf{Figure S12} and DFT Methodology section). Here, the octahedral site intercalation, Mo$_{oct}$ within the van der Waals spacing results as thermodynamically most favourable by 23\% lower in energy of formation than other interstitial at tetrahedral sites, Mo$_{tet}$ in the host material 2H-MoS$_2$ (see \textbf{Table S7} in SI). This is also in agreement with earlier DFT studies of the Mo impurities in the bi-layers of the 2H-MoS$_2$, with a possible spin-centre at the Mo$_{oct}$ sites as favourable outcome\cite{Cortes2018}. We found that the choice Mo$_{oct}$ indeed results a high-spin state due to 4 unpaired electrons aligned parallel at the Mo$^{2+}$(4d$^4$) site as seen from the pDOS analysis presented in \textbf{Figure \ref{fig:MoOct}}. However, the other neighbouring tetrahedral site from the Mo$_{tet}$ intercalation remain at the possible weaker intermediate spin-states with Mo$^{3+}$(4d$^{3}$) occupancies (see also SI \textbf{Figure S17}). In the \textbf{Figure \ref{fig:MoOct}} panel a) and b) total and the Mo$^{2+}$(4d$^4$) pDOS is shown with its orbital decomposition, respectively. While in the top panel DOS plots, total DOS of the 2H-MoS$_{2-x}$ is presented with black shaded area, along with other total DOS of Mo$^{4+}$(4d$^2$), S(3p) and Mo$^{2+}$(4d$^4$), but in the lower panel DOS plots, the non-degenerate Mo-4d orbitals of Mo$^{2+}$(4d$^4$) are shown with color lines: Mo-d$_{yz}$/d$_{xz}$ in red, Mo-d$_{z^2}$ in deep green and Mo-d$_{xy}$/d$_{x^2 - y^2}$ in orange/blue dashed or solid lines. Corresponding spin-density plots at Mo-site with induced spin-centres Mo$^{2+}$(4d$^4$) and neighbouring sites are shown in the panels See \textbf{Figure \ref{fig:MoOct} c)} and \textbf{d)} with side and top views, respectively. A schematic sketch on their possible 4d-orbitals occupancies is zoomed out at right side panels. Stronger charge-density with calculated net absolute spin-magnetic moments 3.67$\mu_B$ at  Mo$_{oct}$ sites is the possible signature of the assigned high-spin state of from the Mo$^{2+}$(4d$^{4}$) sites, thus reasonable to the current observed spins of the V1 centre from EPR analysis. 

Indeed, the calculated spin-magnetic moment from the current PBE-D3+U calculations are 3.67$\mu_B$ and 3.34$\mu_B$ at the Mo$_{oct}$ and Mo$_{tet}$ sites, respectively in these S-deficit models. From a general expectation with formal octahedral and tetrahedral crystal field splitting of the 4d-orbitals of MoS$_{6}$ and MoS$_{4}$ interstitial motifs, respectively, resulted to be high-spin (S = $\frac{4}{2}$) and intermediate spin-states (S = $\frac{3}{2}$) and validates presence of the so called V1 charge-centres observed from current EPR analysis, thus corroborated with DFT calculated data regarding its origin search. Hence, the combined impact of the Mo$_{oct}$ and Mo$_{tet}$ models might forms the possible spin-centres spin-$\frac{4}{2}$ and spin-$\frac{3}{2}$ as expected with the Rabi spin-relaxation measurements (see Exp. Sections). Despite lack of first-principles values, the fine details of the $g$-tensor anisotropy observed in the current CW-EPR data is also supported with shape of the DFT calculated spin-densities. Very weaker binding of these V1 sites (77\% weaker energy vs. V2) compared to the native V2 sites, the localized spin-centres at V1 in 2H-MoS$_{2-x}$, are mainly resulted from Mo interstitial during hydrothermal synthesis. It is also noted that due to the additional covalencies with the neighbouring S-atoms with Mo$_{oct}$ sites or also in tetrahedral sites, Mo$_{tet}$ the NN sub-layers of the host 2H-MoS$_{2-x}$ got polarized with extra electrons and looks honeycomb polaronic region and possible source of polaritons in such 2D nano-structures, see also \textbf{Figure \ref{fig:MoOct}c)} and \textbf{d)}\cite{Mili2015} and SI \textbf{Figure S17 c)} and \textbf{d)}. 

To further recheck the robustness of the present strain imposed defect sites engineering via DFT modelling of the S-vacancy and Mo-interstitials within the 2H-MoS$_{2-x}$ models, we have also calculated their spin properties taken into account the nano-size 2D-slabs of the host material. These earlier results essentially remain in the same trend, qualitatively and quantitatively. As discussed the V1 sites from the Mo$_{oct}$ leaves about 4 unpaired parallel electrons located in the 4d-orbitals resulted Mo$^{2+}$ charge centre and V2 sites result similarly Mo$^{3+}$ charge-centre due to one unpaired parallel electrons. Details of the spin-density plots are presented in the SI \textbf{Figure S19} with side and top views. Also their pDOS analysis of the Mo$^{2+}$(4d$^4$) orbitals are quite reasonable compared to their spin-density plots (See SI \textbf{Figure S20} and \textbf{Figure S21} ). The calculated spin-magnetic moments are 3.9 $\mu_B$ and 1.4 $\mu_B$ respectively for the V1 and V2 defect sites. A partial enhanced spin-density value may be due to the extra change-density mediated from the slab-vacuum model using the present DFT defects sites engineering.

\section{Conclusion}
In summary, we have investigated the origin, spin-states and dynamics of the localised spin-centres in the hydrothermally synthesised 2H-MoS$_{2-x}$ nanocrystals. The 2-6\% in-plane and 2-4\% out-of-plane tensile strains estimated from the first principle calculations were found to play a critical role behind the spin localisation around the Mo$^{2+}$ and Mo$^{3+}$ charge-centres in our 2H-MoS$_{2-x}$ nanocrystals, rather than the spin-neutral Mo$^{4+}$ centres in standard 2H-MoS$_2$. These localised charges appeared as paramagnetic spin centres in the material, namely molybdenum interstitial (V1) and sulfur vacancy (V2) as identified by the EPR measurements and corroborated by DFT calculations. A molybdenum interstitial atom residing in the octahedral (Mo$^{4+}$) or tetrahedral (Mo$^{3+}$) van der Waals spacing of 2H-MoS$_{2-x}$ nanocrystals can be identified as the origin of V1 centre. The net spin magnetic moment of the octahedral and tetrahedral sites has come out as 3.3-3.7$\mu_B$, which indicates a possible spin-$\frac{4}{2}$ to spin-$\frac{3}{2}$ states for the V1 centres as confirmed by the pDOS and spin-density plots from PBE-D3+U calculations. A sulfur vacancy resulted into three mono-electron Mo$^{3+}$ electron-polaronic centres (situated within 2.5\AA\space) with the spin-magnetic moment of 0.99$\mu_B$ per site. These three Mo$^{3+}$ centres couple through direct magnetic-exchange to form a spin-$\frac{3}{2}$ spin-centre, which is the origin of the V2 centre. The possibility of high spin states for V1 and V2 centres is also supported by Rabi nutation and EDNMR measurements. On the other hand, oxygen vacancy of oxysulfide phase can be identified as V3 centre, which resides in the spin-$\frac{1}{2}$ state as supported by Rabi nutation and literature reports. 

We have also shed light on the evolution of defects as well as on their spin dynamics with varying growth condition of the 2H-MoS$_{2-x}$ nanocrystals. Sulfur-deficit growth condition leads to the maximization of the V3 centre (sample A), while an extreme sulfur-rich growth condition leads to the maximization of V1 defect (sample E). In the intermediate regime, V2 centre appears in majority (sample B, C and D). Also, it has been observed that the vacancy defects (V2 and V3 centres) are relatively slow relaxing spin centres following Raman-like spin-lattice relaxation mechanism as compared to the interstitial defects (V1), which follow direct-like spin-lattice relaxation mechanism at low temperatures ($\textless$ 85K). The appearance of high spin defects with the known spin-relaxation properties may induce ferromagnetic ordering in nano-structured 2H-MoS$_{2-x}$, which will help the future research in the spin-based device applications or so called quantum technology based devices with lower energy consumption.

\section{Experimental Details}
\label{sec.exp}
\subsection{Materials Preparation and Characterization}
MoS$_{2-x}$ nano-crystalline samples, with varying S/Mo atomic ratio, were synthesised through a typical hydrothermal method (synthesis temperature: 200$^\circ$C, synthesis time: 24 hr). In the synthesis, 20 mL of  thiourea solution (dissolved in DI water) was added drop-wise to the 15 mL of 0.1 M aqueous MoO$_3$ solution. Both the precursors were procured from Sigma-Aldrich, India and were used without further processing. This total solution was stirred vigorously for 1 hr to make a homogeneous mixture and was transferred to a 60 mL teflon-lined stainless steel autoclave. After reaction, the precipitate were cleaned several times with water and ethanol followed by heating at 60$^\circ$C to obtain  2H-MoS$_{2-x}$ nanocrystals. The S/Mo atomic ratio of the 2H-MoS$_{2-x}$ nanocrystals was altered through the reaction of MoO$_3$ and thiourea with varying molar ratio. We synthesised five number of as-synthesised samples, namely A, B, C, D and E, where the precursors were taken in the ratio of 1:1, 1:2, 1:5, 1:8 and 1:10, respectively. Apart from the as-synthesied form, sample C (1:5 molar ratio) was further annealed in argon environment for 3 hr, and was named as sample C1 in the measurements.

XRD measurements were performed in Bruker D2 phases table top spectrometer. XPS measurements were carried out by an ULVAC-PHI 5000 Versa Probe II spectrometer. Raman measurements were executed by a T64000 (JY, France) micro-Raman spectrometer. X-band EPR measurements were performed in a Bruker ELEXYS E-580 spectrometer coupled with helium cooled closed cycle cryo-system. W-band EPR experiments were conducted in a home-built spectrometer developed at Goldfarb lab, Weizmann Institute of Science, Israel.

\section{Theory and Methodology in DFT Modelling}
\label{sec:dft2}
\subsection{Defects Modeling Using DFT}
Two major challenges have been mimicked through the approximated manner in the current DFT analysis of the point-defects at the moderate computational resource requirement and reasonable accuracy. First, the true nano-structure environment of the 2H-MoS$_2$, is modelled through the applied external strain on the bulk phase, because of the observed nano-particles size in the current exp. synthesis are comparable with bulk properties in 2D and found to be validated in our earlier work on quantum-confinement in layered dichalcogenides \cite{das2022quantum}. And secondly, the choice of the DFT exchange-correlation is set to PBE-D3+U formulation through the systematic search of the on-site Hubbard U either on Mo(4d) and or S(3p). The calculated data is bench marked with earlier single sulfur vacancy formation energies, S$_v$  as reported from the higher level of post-DFT theory and modelling and also the use of the PBE-D3+U method is motivated from similar works on the layered oxides V$_2$O$_5$ and MoO$_3$ while one looks into the thermochemistry and electronic band gaps at the moderate computational loads \cite{das2019structural,das2021layered}. These point defects in the current model of nano-structured 2H-MoS$_2$ is opted within a supercell of size 3$\times$3$\times$1, while the strain generated in the exp. synthesized sample was induced along with various external stimuli through in-plane tensile strain, i.e. 2.0\%, 4.0\%, 6.0\% and a combined 6.0\% in-plane and 4.0\% out-of-plane tensile strain on the first-principles volume of the bulk 2H-MoS$_2$. The highest strain essentially correlated with observed exp. lattice parameter of the nano-structured 2H-MoS$_2$ in the current measurements, yielding values a = b = 3.161$\AA$ and c= 12.965$\AA$ (see also Table \ref{tbl:table1}), compromising lattice value of 2H-MoS$_{2-x}$ than the standard sample 2H-MoS$_{2}$. The single sulfur vacancy S$_v$ formation energy (E$_{form}$) is estimated from the current first-principles theoretical calculations at absolute 0K with Eq. \ref{eq:2}
\begin{equation}
\label{eq:2}
E_{form}(S_v) = \left(E_{sv}-E_{pure}+\mu_s\right)
\end{equation}
Where, E$_{pure}$ is the total energy of the pure 2H-MoS$_2$ supercell without sulfur vacancy (pristine sample) and $E_{Sv}$ is the total energy from the single $S_v$ included 2H-MoS$_2$ supercell model and $\mu_S$ is the chemical potential of the sulfur at the Mo-rich/S-poor condition at the thermal equilibrium as defined with Eq. \ref{eq:3},
\begin{equation}
\label{eq:3}
\mu_s = \frac{1}{2}\left(\mu_{MoS_2}-\mu_{Mo(bcc)}\right)
\end{equation}
Here, $\mu_{MoS_2}$ is the total energy of the bulk 2H-MoS$_2$ unit cell and $\mu_{Mo(bcc)}$is the total energy of the known stable bcc cubic (body-centred cubic) unit cell of Mo, and the strategy has been well accepted in the earlier DFT studies and also values found remarkably well agreed with earlier DFT literature values performed at higher DFT Jacob's Laddar rung\cite{dolui2013possible,tan2020stability}. While for the other two defect with Mo$_{tet}$ and Mo$_{oct}$ the formation energy at the same footing of the current DFT formulation is estimated with the following Eq. \ref{eq:4},
\begin{equation}
\label{eq:4}
E_{form}(Mo_{tet} \hspace{1mm} or \hspace{1mm} Mo_{oct}) = \left(E_{Mo}-E_{pure}-\frac{1}{2}\mu_{Mo(bcc)}\right)
\end{equation}
Here $E_{Mo}$ is the total energy of the 2H-MoS$_2$ bulk supercell containing the Mo intestinal defects (Mo$_{tet}$  or Mo$_{oct}$) i.e. Mo atom is placed at the tetrahedral or octahedral site in the van der Waals spacing of the host lattice. A schematic diagram of these defect models is presented in the SI \textbf{Figure S12}. Only, upon the strained conditions these three charge centres are likely to be most relevant in the current context as it is also consistent with the current exp. observation, while the other un-strained or uniaxial strained or even with V$_{SO}$ centres (oxygen atoms at S$_v$ vacancy), these charge centres are invisible and to be discussed briefly in main text. The calculated defect formation energies of the Mo-interstitial are presented in \textbf{Table S6} in the SI. \\
All first-principles calculations within density functional theory (DFT) for the crystal structure optimization with/out volume relaxation were done using PAW (projector Augmented wave)\cite{paw1,paw2} and Ultrasoft pseudo-potentials (USPP) for the Mo and S atoms, separating the core and valence electrons from the SSSP database as implemented with the DFT software package Quantum Espresso version 6.8\cite{Giannozzi_2009,Giannozzi_2017,Giannozzi_2020}. In the pseudo potential choice, we have 14 valence electrons for Mo atoms, [Kr]4s$^2$4p$^6$4d$^5$5s$^1$ and 6 valence electrons for the S atoms with configuration [Ne]3s$^2$3p$^4$. A kinetic energy cut-off of 75 Ry and a cut-off for density 700 Ry was sufficient to get well-conversed structure of the bulk 2H-MoS$_2$ and $S_v$ vacancy energies through the spin-polarized calculations. The required convergence was achieved within a 3$\times$3$\times$1 supercell (total 54 atoms in the supercell model), along with convergence criteria for total energy threshold up to 1.0D-12 Ry and force up to 1.0D-05 a.u. A Monkhorst-Pack (MP) type \cite{monkhorst1976special} k-mesh 16$\times$16$\times$7 and 4$\times$4$\times$3, respectively, for the bulk and supercell structure for the Brillouin-zone integration, was used. We have opted standard BFGS algorithm\cite{marzari1999thermal} in QE6.8 for the structural relaxations (either for lattice positions or both lattice positions and cell volume relaxation), using the Marzari-Vanderbilt (m-v) smearing technique with a broadening parameter 0.01 Ry\cite{cococcioni2005linear}. \\
All spin-polarized ground state calculations on the defect included models are done to get initial ground state density from the in-built generalized gradient approximation (GGA) as proposed from Perdew-Bruke-Ernzerhof (PBE)\cite{perdew1996generalized} implemented within the pseudopotentials, along with additional onsite Hubbard $U$-term for Mo(4d) and S(3p) from the proposition from A. I. Liechtenstein and V. Anisimov\cite{anisimov1991band,anisimov1993density,anisimov1997first,liechtenstein1995density}, and dispersion energy correction at the D3-level\cite{grimme2010consistent,grimme2011effect} i.e. the so-called PBE-D3+U exchange-correlation functional. As mentioned earlier, the choice of the DFT+U exchange-correlation to take into account the partial d-electron occupancies is a better ways, and have become a vital aspect within the layered materials composed with 3d- or 4d-transition metal cations including Ti, V, Mo, W and Ru\cite{Chakraborty2017,das2019structural,das2023honeycomb}, materials with a distinct class with van der Waals spacing within the periodic unit cell. \\
Thus, in the current work, a systematic search of the on-site Hubbard U term on the Mo- and S-sublattices are done while reference to their exp. band gap ($\sim$ 1.3 eV indirect) and expenses of S$_v$ formation was tested systematically in the nano-structured 2H-MoS$_2$. In the SI \textbf{Figure S13 and Figure S14}, we have shown the calculated electronic band-edges position calculated using different Hubbard onsite-$U$ values applied either on the Mo(4d) or on both Mo(4d) and S(3p) orbitals. The calculated band gap looks reasonable while one include Hubbard effective $U$ = 4 eV both on the Mo(4d) and S(3p) with calculated indirect band gap at the DFT level 0.85 eV from the first-principles relaxed volume of 2H-MoS$_2$ bulk unit cell (see also SI \textbf{Table S5}). Compared to the exp. value, though this is about 35\% lower band gap vs. exp. values, but better than than the standard DFT data 0.73 eV (about 50\% underestimated). \\ 
Finally, we notice also the vacancy formation energies of the $S_v$ at the given exchange-correlation with or without $U$ term (see \textbf{Table S6} in SI), and found consistent result than earlier reported values, and proved the robustness of the current calculations strategy\cite{zheng2014tuning,tan2020stability}. Nonetheless, the most demanding aspect of the applying $U$-term on the S-site is actually more interesting and would be disused for 3 major defect models, i.e. sulfur vacancies ($S_v$), Mo-interstitials (tetrahedral site, Mo$_{tet}$ or octahedral site, Mo$_{oct}$) and Oxy-sulfides (V$_{SO}$) at the $S_v$ centres. \\
Finally, these 3 major charge-centres as observed with the QE6.8 software based calculations and simulations of defects are also reproduced with the help the plane-wave and PAW pseudopotential in VASP5.4.4 DFT software package\cite{Kresse1996,Kresse1999} and presented in the main text while other plots from QE6.8 are shown in SI data. The choice of the PBE-D3+$U$ exchange-correlation, valence configuration of the Mo and S atoms, the MP k-mesh grid are kept same footing as it is with the QE6.8 input defination, while a kinetic energy cut-off 600 eV is used in VASP calculation with energy convergence 10D-09 eV and force convergence on the atoms 5 meV/$\AA$. From the single-point runs, the total and atom decomposed density of states are performed with energy resolution 0.01 eV within the Tetrahedron smearing and Gaussian broadening 0.1 eV at the suitable dense FFT grid as we know from earlier DFT simulation studies\cite{das2021layered,das2023honeycomb}. All crystal structure visualization are done in two software using the XcrySDen\cite{kokalj1999xcrysden} and VESTA tool\cite{momma2011vesta}.

\begin{acknowledgement}
S.K. and K.B. acknowledge Ministry of Human Resource and Development (MHRD), India for providing research fellowship. T.D. acknowledge and thank to the research grant for Ramanujan Fellow provided by Science and Engineering Research Board (SERB), Govt. of India, File No. RJF/2021/000120. T.D. also acknowledge National Supercomputing Mission (NSM) for providing computing resources of "PARAM Shakti" at IIT Kharagpur, which is implemented by C-DAC and supported by the Ministry of Electronics and Information Technology (MeitY) and Department of Science and Technology (DST), Government of India. D. B. acknowledge the support of Mr. Rudheer Bapat of TIFR Mumbai for the TEM measurements. All authors acknowledge Central Research Facility (CRF), IIT kharagpur for XRD, XPS, Raman and EPR measurements. Authors also happy to acknowledge the productive discussion with Prof. Gianfranco Pacchioni, Univ. of Milano-Bicocca, Italy.
\end{acknowledgement}

\begin{suppinfo}

All data are available from the corresponding authors upon reasonable request.

\end{suppinfo}

\bibliography{mybiblio}

\end{document}